\documentclass[prl,10pt,twocolumn,preprintnumbers,amsmath,amssymb,superscriptaddress,noshowpacs,nofootinbib]{revtex4-1}
\usepackage{mathrsfs,amsmath,amssymb,amsthm,amsfonts,tikz,graphicx,accents,hyperref, color}
\usepackage{dsfont,epiolmec, latexsym, stmaryrd}
\usepackage{enumerate, comment}
\usepackage{setspace}
\usepackage{footmisc}
\usepackage{mathrsfs, calligra}
\usepackage{multirow}
\usepackage{xcolor}

\hypersetup{ linktoc=all,
    colorlinks, linkcolor={palatinateblue},
    citecolor={brightpink}, urlcolor={amaranth}
}

\graphicspath{{Images/}}

\definecolor{darkgreen}{rgb}{0.1,0.6,0.1}
\definecolor{darkblue}{rgb}{0,0,0.3}
\definecolor{darkred}{rgb}{0.7,0,0}

\definecolor{light gray}{RGB}{220,220,220}
\definecolor{dark purple}{RGB}{108,0,217}
\definecolor{pink}{RGB}{190,20,100}
\definecolor{orang}{RGB}{193,63,0}
\definecolor{green}{RGB}{11,98,17}
\definecolor{darkpink}{RGB}{153,0,76}
\definecolor{bluegreen}{RGB}{0,102,102}
\definecolor{greenlagan}{RGB}{0,102,0}
\definecolor{redgreen}{RGB}{102,102,0}
\definecolor{Redgreen}{RGB}{153,76,0}
\definecolor{vividviolet}{rgb}{0.62, 0.0, 1.0}
\definecolor{amaranth}{rgb}{0.9, 0.17, 0.31}
\definecolor{palatinateblue}{rgb}{0.15, 0.23, 0.89}
\definecolor{brightpink}{rgb}{1.0, 0.0, 0.5}
\definecolor{cornflowerblue}{rgb}{0.39, 0.58, 0.93}
\definecolor{deepcarminepink}{rgb}{0.94, 0.19, 0.22}
\definecolor{radicalred}{rgb}{1.0, 0.21, 0.37}
\def\GN{G_{_{\text{{N}}}}}
\def\rh{r_{_{\text{{H}}}}}
\def\th{T_{_{\text{{BH}}}}}

\def\xih{\xi_{_{\text{{H}}}}}
\def\zetah{\zeta_{_{\text{{H}}}}}
\def\Sbh{S_{\text{\tiny{BH}}}}
\def\SbhW{S_{\text{\tiny{BH}}}^{\text{\tiny{W}}}}

\def\bx{\mathcal{X}}
\def\cG{{\cal G}}
\begin{document}

{\vskip .2cm}
\title{On Black Hole Temperature in Horndeski Gravity}

\author{K.~Hajian}
\email{khajian@metu.edu.tr}
%\affiliation{School of Physics, Institute for Research in Fundamental Sciences (IPM),  P.O.Box 19395-5531, Tehran, Iran}
\affiliation{Department of Physics, Middle East Technical University, 06800, Ankara, Turkey}
\affiliation{Institute of Theoretical Physics and Riemann Center for Geometry and Physics
Leibniz University Hanover, Appelstrasse 2, 30167 Hannover, Germany}

\author{S.~Liberati}
\email{liberati@sissa.it}
\affiliation{SISSA, Via Bonomea 265, 34136 Trieste, Italy and INFN, Sezione di Trieste}
\affiliation{IFPU - Institute for Fundamental Physics of the Universe, Via Beirut 2, 34014 Trieste, Italy}

\author{M.M.~Sheikh-Jabbari}
\email{jabbari@theory.ipm.ac.ir}
\affiliation{School of Physics, Institute for Research in Fundamental Sciences (IPM),  P.O.Box 19395-5531, Tehran, Iran}
\affiliation{The Abdus Salam ICTP, Strada Costiera 11, Trieste, Italy}

\author{M.H. Vahidinia}
\email{vahidinia@iasbs.ac.ir}
\affiliation{Department of Physics, Institute for Advanced Studies in Basic Sciences (IASBS),
P.O. Box 45137-66731, Zanjan, Iran}
\affiliation{School of Physics, Institute for Research in Fundamental Sciences (IPM),  P.O.Box 19395-5531, Tehran, Iran}
%--------------------------------------------------------------------------------------------------------------------------------------
%--------------------------------------------------------------------------------------------------------------------------------------
\begin{abstract}
It has been observed that for black holes in certain family of Horndeski gravity theories Wald's entropy formula does not lead to the correct first law for black hole thermodynamics. For this family of Horndeski theories speeds of propagation of gravitons and photons are in general different and gravitons move on an effective metric different than the one seen by photons. We show that the temperature of the black hole should be modified from surface gravity over $2\pi$ to include effects of this effective metric. The modified temperature, with the entropy unambiguously computed by the solution phase space method,  yields the correct first law. Our results  have far reaching implications for the Hawking radiation and species problem, going beyond the Horndeski theories. 
\end{abstract}


\maketitle

Theories of beyond Einstein general relativity, with either theoretical \cite{Burgess-review} or dark energy or dark matter and cosmological model building motivations \cite{Clifton:2011jh, Nojiri, Creminelli:2018xsv, Sakstein:2017xjx, Ezquiaga:2017ekz, Heisenberg:2018vsk, Barack:2018yly}, have been very extensively discussed in the literature. These theories while generally covariant, typically have fourth order field equations and  hence have propagating ghost degrees of freedom and are pathological. There are, nonetheless, special classes of theories which are ghost free, like the so-called $f(R)$ theories \cite{f(R)-review},  Lovelock theories \cite{Lovelock} and the class of scalar-tensor theories
first formulated and classified by Horndeski \cite{Horndeski:1974wa}. In the last two decades, Horndeski family has also been extended further 
%to the  beyond-Horndeski theories 
\cite{Beyond-Horndeski-1,Beyond-Horndeski-2, Beyond-Horndeski-3, Beyond-Horndeski-4, Beyond-Horndeski-5}.

Black holes are ubiquitous solutions to generally covariant gravity theories, Einstein gravity and beyond, and recent observations of gravity-waves \cite{Abbott:2016blz,LIGOScientific:2018mvr, Akiyama:2019cqa} have put them at forefront of fundamental physics research \cite{Creminelli:2018xsv, Sakstein:2017xjx, Ezquiaga:2017ekz}. Theoretically, black hole are typically specified by having an event horizon,  yet to be confirmed observationally. At the theoretical level, once quantum effects are also taken into account, black holes behave as a thermodynamical system with an entropy  and temperature associated with the horizon and satisfy laws of thermodynamics \cite{Bardeen:1973gd}.

In the ``standard picture'', the  temperature $T_{_\text{H}}$ for black holes with a Killing horizon, is given by the surface gravity $\kappa$ at the horizon as $T_{_\text{H}}=\kappa/(2\pi)$. This  is the same  temperature as the black body radiation  emitted by the black hole,  Hawking radiation \cite{Hawking:1974sw}. The black hole entropy $S_{\text{\tiny{BH}}}$, on the other hand, for Einstein gravity theory is given by Bekenstein-Hawking area law \cite{Bekenstein:1973ur}, $S_{\text{\tiny{BH}}}=A_{_\text{H}}/(4\GN)$, where $A_{_\text{H}}$ is area of horizon and $\GN$ is the Newton constant. The other black hole charges appearing in the first law of black hole thermodynamics, like mass, angular momentum and the electric charge, are then typically computed in the asymptotic region and defined e.g. by the ADM method or its extensions \cite{Arnowitt:1960es,Arnowitt:1962hi}.

Thermodynamic description of black holes is quite universal and applies also to black holes in beyond Einstein gravity; as shown in  two seminal papers \cite{Wald:1993nt,Iyer:1994ys} they are a result of general invariance of the theory. Black hole entropy, however, depends on the theory and in general is not given by the area law. Despite its elegance and very wide success, it has been observed that Wald's entropy formula \cite{Wald:1993nt} and/or the first law does not work for a family of solutions to certain Horndeski theories, e.g. see \cite{Feng:2015wvb}. This is
the problem we will address in this work.

A similar  violation of the first law was reported for some   black holes in Einstein-dilaton theories which have a scalar field $\phi$ with shift symmetry  \cite{Gibbons:1996af} and a similar suggestion was put forward: One may  introduce  a new term in the first law, associating a chemical potential and an ad hoc conserved charge to the scalar field \cite{Feng:2015wvb,Gibbons:1996af}. This proposal, while fixing the issue with the first law, has the problem that the (Noether) conserved charge associated with the scalar shift symmetry  is zero and the charge associated with the scalar is ``ad hoc''. Moreover, for  Einstein-dilaton theory black holes, this
is in contrast with the no-hair theorems and the absence of independent conserved charges associated with such scalar fields. Indeed, this proposal was  refuted by proving that the dilaton moduli are redundant parameters  and that there cannot be a term associated with variation of asymptotic value scalar fields in the first law \cite{Hajian:2016iyp}. 

To tackle the problem with Horndeski black holes we carefully revisit Wald's derivation of the first law and his definition of the entropy. Wald's entropy is based on  the standard Noether method and postulates  $\kappa/(2\pi)$ as the black hole temperature.  Noether's theorem and Wald formula have ambiguities which should be carefully dealt with. As we will briefly discuss, these ambiguities do not all vanish in the case of Horndeski theories. Fortunately, there is another method for calculation of charges (called covariant formulation \cite{Ashtekar:1987hia,Ashtekar:1990gc,Lee:1990gr}) which is free of the ambiguity in Wald's formula. To put this method into its full computational power, we use its version introduced in \cite{Barnich-Compere, Hajian:2015xlp}, the solution phase space method (SPSM).

To read the entropy in SPSM we need to provide the surface gravity and/or the black hole temperature. The key point of the current Letter comes from the fact that in Horndeski theories gravitons do not move with the speed of light \cite{Ezquiaga:2017ekz, Bettoni:2016mij}; they effectively propagate on a spacetime which is especially different than the black hole metric close to the horizon. Therefore, they feel a different surface gravity and hence a temperature different than the usual Hawking temperature $\kappa/(2\pi)$.
This modified temperature, together with the correspondingly defined entropy, results in the correct first law.

We first introduce  Horndeski theories and  the formulae for speed of gravitons in them and show why Wald entropy does not in general work for Horndeski black holes. We then  very briefly review the solution phase space method (SPSM) \cite{Hajian:2015xlp} (which is based on covariant phase space formulation of charges \cite{Ashtekar:1987hia,Ashtekar:1990gc,Lee:1990gr}) for computing the entropy in generally covariant theories and apply it to Horndeski black holes. The SPSM takes surface gravity for gravitons as an input to compute the entropy. We provide this input through effective near horizon metric as seen by gravitons and verify in two examples how our modified temperature restores the first law for the Horndeski black holes. Finally, we discuss the deep implications  our analysis  and results can have for better understanding of Hawking radiation and black hole dynamics.  \\

\centerline{\textbf{ Review of  Horndeski Gravity}}\vskip 2mm 

Horndeski theories are a class of scalar-tensor theories with the action \cite{Horndeski:1974wa, Maselli:2015yva, Gleyzes:2013ooa, Beyond-Horndeski-5, Kovacs:2020ywu}
\begin{equation}\label{Horn-action}
S_{\text{Horn.}}=\frac{1}{16\pi \GN}\int d^nx\sqrt{-g}\ {\cal L}_{\text{Horn.}}
\end{equation}
\begin{equation}\label{Horn-Lagrangian-1}
\begin{split}
\hspace*{-2mm}&{\cal L}_{\text{Horn.}}=\cG_2(\phi,\bx)-\cG_3(\phi,\bx)\Box\phi+\cG_4(\phi,\bx)R \cr
&+\cG'_4(\phi,\bx) \left((\Box\phi)^2-(\partial_{\mu\nu}\phi)^2\right)  
- \cG_5(\phi,\bx) G^{\mu\nu}\partial_{\mu\nu}\phi\cr &\quad-\frac{\cG_5'(\phi,\bx)}{6}\left((\Box\phi)^3+2(\partial_{\mu\nu}\phi)^3-3\Box\phi(\partial_{\mu\nu}\phi)^2 \right)
\end{split}
\end{equation}
where $g_{\mu\nu}$ is the spacetime metric, $R$ is Ricci scalar, $G_{\mu\nu}$ is the Einstein tensor, $\partial_{\mu\nu}\phi=\nabla_\mu\nabla_\nu\phi,\, \,  \Box\phi\!\!=\!\!g^{\mu\nu}\partial_{\mu\nu}\phi$,\, \,  $\bx:=\!-\frac12\partial_\mu\phi\partial^\mu\phi$ and $\cG_i'=d\cG_i/d\bx$. We are adopting the conventions that $\cG_4(\phi=0,\bx=0)=1$. This is how we define the Newton constant $\GN$.

For our analysis below  we restrict ourselves to a large class of models with $\cG_4, \cG_5$ whose Lagrangian 
up to some total derivatives, takes the form \cite{Gleyzes:2013ooa}
\begin{equation}\label{Horn-Lag}
{\cal L}_{\text{Horn.}}
=\cG_2 + (\cG-\cG' \bx) R +\cG'G^{\mu\nu}\partial_\mu\phi\partial_\nu\phi 
\end{equation}
%where $\cG= \cG_4+\bx d\cG_5/d\phi$, $\tilde\cG_2=\cG_2+2d^3\cG_5/d\phi^3, \tilde\cG_3=\cG_3+3d^2\cG_5/d\phi^2$. 
In our analysis below we  assume $\cG'\neq 0$. For $\cG'=0$ cases we recover the usual Brans-Dicke type theory which is not the subject of our analysis here, as Wald entropy formula works for them properly.

Let us consider the  ``$\phi+3$'' decomposition of the Horndeski Lagrangian \eqref{Horn-Lag} along the constant $\phi$ surfaces by taking 
\begin{equation}\label{g-h-n}
    g_{\mu\nu}=h_{\mu\nu}+ \sigma \phi_\mu \phi_\nu, \qquad \phi_\mu:=\frac{\partial_\mu \phi}{|\partial\phi|},
\end{equation}
$\sigma$ is sign of $\phi_\mu \phi^\mu$, it is $-1$ for cosmological backgrounds and $+1$ for black holes, and $h_{\mu\nu}$ is the metric along constant $\phi$ surface, $h_{\mu\nu}\phi^\nu=0$.
The details of the analysis may be found in \cite{Gleyzes:2013ooa} and the result for the ``$\phi+3$'' decomposed Lagrangian is 
\begin{align}\label{Horndeski-frame-decomposed}
\mathcal{L}&= \cG_2+ \cG\ ^{(3)}\!\!R+(\cG-2\bx\cG') ( K_{\mu\nu}K^{\mu\nu}-K^2)\cr &+2\sqrt{-2\bx} \cG_{,\phi} K +\text{total derivative terms}
\end{align}
where $^{(3)}\!\!R$ is the scalar curvature of $h_{\mu\nu}$, $K$ is the extrinsic curvature of our constant $\phi$ surfaces, $K_{\mu\nu}=h_\mu^\alpha \nabla_\alpha \phi_\nu,\ K=K^\mu{}_\mu$ and $\cG_{,\phi}=d\cG/d\phi$.

\noindent\textbf{Speed of gravitons on black hole backgrounds.} 
%Gravitons in Horndeski theory move with a speed different than speed of photons. 
To  compute the speed of gravitons, one should systematically study linearized field equations around a given background, in our case a black hole. While this can be done, see e.g. \cite{Kobayashi:2012kh, Kobayashi:2014wsa},  the above ``$\phi+3$'' decomposition provides a shortcut.

For a black hole $\phi_\mu$ is typically along the ``radial direction'' and is  normal to the horizon and,  the time direction  is in the ``3'' part,  along $h_{\mu\nu}$ metric and normal to $\phi_\mu$.
From \eqref{Horndeski-frame-decomposed} one may then directly read the speed of gravitons which  is now  direction dependent and for the case of black holes \footnote{Note that in most of the Horndeski literature which deals with cosmological background, $\phi_\mu$ is timelike. For this case \eqref{Horndeski-frame-decomposed} still holds but then leads to $c_g^2=\frac{\cG}{\cG-2\bx\cG'}$ for all gravitons \cite{Bettoni:2016mij}.} is \cite{Kobayashi:2014wsa}:
\begin{equation}\label{speed-of-graviton}
   \hspace*{-4mm} c_g^2=\begin{cases}\frac{\cG-2\bx\cG'}{\cG}& \text{for gravitons moving along}\ \phi_\mu\\ \frac{\cG}{\cG}=1 &\text{for gravitons moving normal to}\ \phi_\mu\end{cases}
\end{equation}

\vskip 2mm

{\centerline{\textbf{Wald entropy formula and Horndeski theory}}} 
\vskip 2mm
Consider a covariant gravitational theory described by the Lagrangian ${\cal L}={\cal L}(g_{\mu\nu}, R_{\mu\nu\alpha\beta},\nabla_\rho R_{\mu\nu\alpha\beta},\cdots)$, where $g_{\mu\nu}$ is the spacetime metric, which we take to be $n$ dimensional, $R_{\mu\nu\alpha\beta}$ is its Riemann curvature and $\nabla_\rho$ is its covariant derivative. Horndeski theory \eqref{Horn-Lag} is an example of such theories. The Wald entropy for a black hole solution to this theory is defined as \cite{Wald:1993nt}
\begin{equation}\label{SWald}
\SbhW:= 2\pi \int_{\text{H}}\mathbf{X}^{\mu\nu}\boldsymbol{\epsilon}_{\mu\nu},
\end{equation}
in which $\boldsymbol{\epsilon}_{\mu\nu}$ is the binormal tensor to the ($n-2$)-surface H associated to the black hole horizon, normalized as $\boldsymbol{\epsilon}_{\mu\nu}\boldsymbol{\epsilon}^{\mu\nu}=-2$, and satisfying the identity 
\begin{equation}\label{binormal}
(d\xi_{_\text{H}})_{\mu\nu}={2}\kappa \boldsymbol{\epsilon}_{\mu\nu},
\end{equation}
where $\xih$ is the horizon Killing vector and $\kappa$ is surface gravity of the black hole, and $\mathbf{X}^{\mu\nu}$  is  
\begin{equation}\label{X form}
(\mathbf{X}^{\mu\nu})_{\mu_3 \dots  \mu_n}=- \frac{\delta \mathcal{L}}{\delta R_{\alpha\beta\mu\nu}}\boldsymbol{\epsilon}_{\alpha\beta\mu_3\dots\mu_n},
\end{equation}
where $\boldsymbol{\epsilon}_{\mu_1\mu_2\mu_3\dots\mu_n}$ is the spacetime volume form \cite{Wald:1993nt,Iyer:1994ys}. 

The Wald entropy formula has been extremely successful in providing the correct entropy in black hole literature,  nonetheless it suffers from an ambiguity which can yield wrong entropy in special cases. As we show below Horndeski theories are among these special cases. The Horndeski Lagrangian \eqref{Horn-Lag} has a $G^{\mu\nu}\partial_\mu\phi\partial_\nu\phi$ term. Since $G^{\mu\nu}=R^{\mu\nu}-\frac12 g^{\mu\nu}R$, we have terms like $R^{\mu\nu}\partial_\mu\phi\partial_\nu\phi$, which can be rewritten as
\begin{align}\label{Ricci-identity}
 &R^{\rho\sigma}\partial_\rho\phi\partial_\sigma\phi =-(\nabla_\rho \nabla_\sigma \phi)^2+(\Box\phi)^2+
\nabla_{\mu}{\cal W}^\mu,
\end{align}
with
\begin{align}
& {\cal W}^\mu =(\nabla_\nu\phi\nabla^\nu\nabla^\mu\phi-\Box\phi\nabla^\mu\phi).
\end{align}
The explicit dependence on the Ricci tensor $R_{\mu\nu}$ can hence be removed in favor of terms with $\phi$ derivatives. A careful analysis \cite{Work-in-progress} reveals that this yields an ambiguity in $\mathbf{X}^{\mu\nu}$ \eqref{X form}, a ``$\mathbf{W}$ ambiguity'' in Wald's terminology \cite{Wald:1993nt,Iyer:1994ys},
\begin{equation}\label{X deformed}
(\mathbf{X}^{\mu\nu})_{\mu_3 \dots  \mu_n} \,  \to \,  (\mathbf{X}^{\mu\nu})_{\mu_3 \dots  \mu_n} +\lambda\frac{\delta R^{\rho\sigma}}{\delta R_{\alpha\beta\mu\nu}}\partial_\rho\phi\partial_\sigma\phi\boldsymbol{\epsilon}_{\alpha\beta\mu_3\dots\mu_n}.    
\end{equation}
where $\lambda$ is an arbitrary number. The last term above can be non-zero  and contribute to the Wald formula \eqref{SWald} (see supplemental material for details). 
%So, the definition of Wald entropy in Horndeski theories is ambiguous. 
To circumvent this problem in Horndeski gravity, we use the solution phase space method which is free of this ambiguity. \\

{\centerline{\textbf{Entropy in solution phase space method}}} 
\vskip 2mm

Consider an $n$ dimensional  generally covariant theory described by a Lagrangian ${\cal L}$ and denote the dynamical fields collectively  by $\Phi$ and  its generic solutions by $\bar\Phi$. Let the $n$-form $\mathbf{L}$ be the Hodge dual of the Lagrangian. The Noether  current  $(n-1)$-form $\mathbf{J}$ associated with a smooth vector $\xi^\mu$ is then
\begin{equation}\label{Noether current}
\mathbf{J}_\xi= \mathbf{\Theta}(\delta_\xi \Phi)- \xi\cdot \mathbf{L},
\end{equation}
where $\delta_\xi\Phi$ are Lie-derivative of fields along $\xi$ and the $\mathbf{\Theta}$ term is the standard surface $(n-1)$-form which is read by the variation of the Lagrangian $\delta \mathbf{L}=\mathbf{E}\delta \Phi + d \mathbf{\Theta}(\delta \Phi)$; $d$ denotes exterior derivative on the  space-time, and $\mathbf{E}$ represents the equations of motion. Using the identity $d(\xi\cdot \mathbf{L})=\delta_\xi \mathbf{L}$, then $d\mathbf{J}_\xi=\mathbf{E}\delta_\xi \Phi. 
$ This is the celebrated Noether theorem: by the on-shell condition $\mathbf{E}=0$, $d\mathbf{J}_\xi=0$. As a result, by the Poincar\`{e} Lemma $\mathbf{J}$ is an exact form on-shell, i.e. $\mathbf{J}_\xi=d\mathbf{Q}_\xi$. Noether charge density $\mathbf{Q}$ is an $(n-2)$-form which is locally built out of $\Phi$ and $\xi$.  One can define an $n-2$ dimensional form \cite{Lee:1990gr} 
\begin{equation}
\boldsymbol{k}_\xi(\delta\Phi, \bar\Phi):=\delta \mathbf{Q}_\xi- \xi\cdot \mathbf{\Theta}(\delta \Phi)   
\end{equation}
where  $\delta\Phi$ is a generic variation of the fields satisfying linearized field equations.

The variation of the Hamiltonian generator associated with flows of $\xih$ is given by \cite{Iyer:1994ys,Hajian:2015xlp}
$
\delta H_{\xih}=\int_{\text{H}} \boldsymbol{k}_\xi(\delta\Phi, \bar\Phi).$ The entropy variation $\delta \Sbh$ which satisfies a consistent first law is then defined as $\delta H_{\xih}:=\th \delta\Sbh$, i.e.
\begin{equation}\label{entropy-formula}
    \delta \Sbh:=\frac{1}{\th}\int_{\text{H}} \boldsymbol{k}_{\xih}(\delta\Phi, \bar\Phi),
\end{equation}
where $\th$ is the black hole temperature, which should be  a purely geometric quantity and  a constant over $\text{H}$. %The $\th$ is a function of black hole parameters such that it makes $\delta S$ integrable in SPSM. 
The definition \eqref{entropy-formula} yields an entropy variation free of $\mathbf{W}$ ambiguity \cite{Hajian:2015xlp, Ghodrati:2016vvf,Hajian:2015eha,Seraj:2016cym,Compere-lecture}. 

The key question in this approach is to how fix $\th$.  In usual cases \cite{Wald:1993nt,Iyer:1994ys} $\th=\frac{\kappa}{2\pi}=T_0,$ where $\kappa$ is the horizon surface gravity and $T_0$ is the Hawking temperature \cite{Hawking:1974sw}. Below we identify $\th$ in Horndeski theories.

\begin{center}
{\textbf{Effective Metric for Gravitons (EMG) and effective surface gravity}}
\end{center}

 Given \eqref{speed-of-graviton} and the direction dependence of the speed of gravitons, one may ask what is the  ``effective'' metric ${\mathtt{g}}_{\mu\nu}$ whose null rays, ${\mathtt{g}}^{\mu\nu}k_\mu k_\nu=0$, the gravitons move on. From \eqref{speed-of-graviton} it is easy to write this metric:
\begin{equation}\label{Disformal-map}\begin{split}
{\mathtt{g}}_{\mu\nu} &=(\cG-2\bx \cG') g_{\mu\nu}-\cG' \partial_\mu\phi\partial_\nu\phi \cr 
& =(\cG -2\bx\cG') h_{\mu\nu}+\cG \phi_\mu \phi_\nu
\end{split}
\end{equation}
where $\phi_\mu$ is defined in \eqref{g-h-n} and to avoid having a singular effective metric we assume $\cG', \cG-2\bx\cG'\neq 0$. In  the terminology of Horndeski gravity literature, the above is a ``disformal map'' \cite{Bettoni:2013diz, Ezquiaga:2017ekz, Bettoni:2016mij} from original spacetime metric to the EMG. 

In order to  compute the surface gravity as seen by gravitons using \eqref{Disformal-map}, we note that the horizon generating Killing vector $\xih^\mu$ is  normal to $\phi_\mu$ at the horizon and one may hence use this to compute the surface gravity, 
\begin{equation}\label{dxi-calE-1}
    d\xih ={2}\kappa c_g {\boldsymbol{{\cal E}}}\quad \text{at the horizon},
\end{equation}
where $\kappa$ is the surface gravity in the matter metric $g_{\mu\nu}$, $c^2_g=\frac{\cG-2\bx\cG'}{\cG}$ and ${\boldsymbol{{\cal E}}}$ is the {bi-normal tensor to} the bifurcation surface H, normalized as  ${\boldsymbol{{\cal E}}}_{\mu\nu}{\boldsymbol{{\cal E}}}^{\mu\nu}=-2$. In particular, note that to lower and raise indices on $\xih^\mu, {\boldsymbol{{\cal E}}}$ we should use  ${\mathtt{g}}_{\mu\nu}$ as in \eqref{Disformal-map}. 
%We will show more details of the derivation of \eqref{dxi-calE} in a couple of examples we discuss. 

To relate to the entropy formula \eqref{entropy-formula}, however, we need to rewrite the above in terms of ${\boldsymbol{\epsilon}}$, the volume two form of the original metric $g_{\mu\nu}$. Recalling \eqref{Disformal-map} we have
\begin{equation}\label{dxi-calE-2}
    {\boldsymbol{{\cal E}}}=\sqrt{\cG(\cG-2\bx\cG')}\ {\boldsymbol{\epsilon}}.
\end{equation}
Inserting this into \eqref{dxi-calE-1} we obtain
\begin{equation}\label{dxi-calE}
    d\xih ={2}\kappa (\cG-2\bx\cG') {\boldsymbol{\epsilon}},
\end{equation}
implying  %$\kappa_{_{\text{\tiny{graviton}}}}={\kappa}(\cG-2\bx\cG')$ and hence
\begin{equation}\label{T-desired}
    T_{_{\text{\tiny{graviton}}}}=(\cG-2\bx\cG') T_0,
\end{equation}
where $T_0=\frac{\kappa}{2\pi}$ is the ordinary Hawking temperature and we still use \eqref{entropy-formula} to compute the entropy. 

The key point is, it is only by identifying $T_{_{\text{\tiny{graviton}}}}=\th$ that we get a consistent first law using \eqref{entropy-formula}. That is, the SPSM indicates that the integrable entropy is the charge of the vector,
\begin{equation}\label{zeta-H}
   \zetah:=\frac{1}{T_{_{\text{\tiny{graviton}}}}} \xih = \frac{2\pi}{\kappa \cdot(\cG-2\bx\cG')}\xih.
\end{equation}
We note that the arguments of exponential peeling of graviton null rays  close to the horizon \cite{Peeling-2013}, leads to the same temperature  as \eqref{T-desired}, {once we consider the appropriate scaling of units \eqref{dxi-calE-2}.}

\vskip2mm

\centerline{\textbf{Examples}}\label{sec:5}
\vskip2mm

To show how our modified temperature  resolves the first law issue for Horndeski black holes we discuss two examples investigated in \cite{Cisterna:2014nua,Feng:2015wvb, Miao:2016aol,Hu:2018qsy}. Three more examples have been discussed in the supplemental material. 

\noindent{\textbf{Example 1:}} In our first example, we study a spherically symmetric black hole in the Horndeski gravity,
\begin{equation}\label{Horndeski L}
\mathcal{L}=\frac{1}{16\pi\GN}\Big(R-F_{\mu\nu}F^{\mu\nu}+2\gamma G^{\mu\nu}\partial_\mu\phi\partial_\nu\phi\Big)
\end{equation}
This action corresponds to $\cG_2=0, \cG_4=1, \cG_5=2\gamma\phi$ in \eqref{Horn-Lagrangian-1}, which yields $\cG=1+2\gamma \bx$ in \eqref{Horn-Lag},  and $F=dA$ is the electromagnetic field strength. This theory has a charged black hole solution \cite{Feng:2015wvb}, 
\begin{equation}\label{metric ansatz}
ds^2=-h(r)dt^2+\frac{dr^2}{f(r)}+r^2(d\theta^2+\sin^2\theta d\varphi^2),
\end{equation}
where 
\begin{align}
h=1-\frac{2m}{r}+\frac{q^2}{r^2}-\frac{q^4}{12r^4}, \qquad f=\frac{4r^4h}{(2r^2-q^2)^2}.
\end{align}
The gauge and scalar fields can also be written as
\begin{align}
A= \left(\frac{q}{r}-\frac{q^3}{6r^3}\right)dt, \qquad d\phi=\sqrt{\frac{-q^2}{2\gamma r^2 f}}dr.
\end{align}
To have a real $\phi$ we take $\gamma<0$. This solution is asymptotically flat and horizon is at $h=f=0$, 
\begin{equation}
    \rh-{2m}+\frac{q^2}{\rh}-\frac{q^4}{12\rh^3}=0.
\end{equation}
Note that, while  the derivative of the scalar field diverges at the horizon, $\gamma\partial^\mu\phi\partial_\mu\phi=\frac{-q^2}{2r^2}$  is finite at the horizon.

The standard methods for calculating conserved charges yields $M=\frac{m}{\GN}$ is the mass and $Q=\frac{q}{\GN}$ is the electric charge of the black hole. Moreover, the surface gravity and horizon electric potential are \cite{Feng:2015wvb},
\begin{equation}
\kappa=\frac{2r^2_{_\text H}-q^2}{4r^3_{_\text H}}, \qquad \Phi_{_\text H}=\frac{q}{r_{_\text H}}-\frac{q^3}{6r^3_{_\text H}}.
\end{equation}
With the Hawking temperature $T_0=\frac{\kappa}{2\pi}$ together the Wald entropy, which yields the usual area law for this example $S=\pi \rh^2/\GN$ (see the appendix for more details of the computation), the first law $\delta S= \frac{1}{T_0}(\delta M-\Phi_{_\text H}\delta Q)$ does not hold. Moreover, with $\th=T_0$ the entropy obtained in the SPSM is not even integrable over parameters $m$ and $q$ of the solution \cite{Feng:2015wvb}. This can be easily seen by replacing all terms in RHS of $\delta S$ above in terms of $m$, $q$ and observe that it is not variation of some $S(m,q)$. 

However, the new temperature in \eqref{T-desired},
\begin{equation}\label{ex1 T}\begin{split}
\th &= (\cG-2\bx\cG')\Big|_\text{H}T_0\\
&= \left(1-\frac{q^2}{2r^2_{_\text{H}}}\right)T_0=\frac{{\left(1-\frac{q^2}{2r^2_{_\text{H}}}\right)^2}}{4\pi\rh},
\end{split}
\end{equation}
makes the entropy computed using \eqref{entropy-formula} integrable, which for this example is given by the usual 
Bekenstein-Hawking entropy $\Sbh=\pi \rh^2/\GN$. With this entropy and temperature \eqref{ex1 T}, it is immediate to verify that the first law,
$$
\th\delta \Sbh =\delta M-\Phi_{_\text H}\delta Q
$$
is also satisfied. Note also that $\th\leq T_0$.

\noindent{\textbf{Example 2:}} Recalling \eqref{T-desired}, the class of models with $\cG-2\cG'\bx=1$ seem to be special. That is, for $\cG=1+2\beta \sqrt{-\bx}$ we do not expect a temperature shift. This is in fact confirmed from  the black hole solution discussed in \cite{Babichev:2017guv}. The Lagrangian of the theory is
\begin{equation}
\mathcal{L}=\frac{1}{16\pi \GN}\Big( (1+{\beta}\sqrt{-\bx})R-2\Lambda +\eta \bx -\frac{\beta}{2\sqrt{-\bx}} {\mathbf\Psi}\Big)
\end{equation}
where ${\mathbf\Psi}:=(\Box\phi)^2-(\partial_{\mu\nu}\phi)^2$ and $\beta, \eta$ are constants. We consider the  black hole solution \cite{Babichev:2017guv} with the metric of the form \eqref{metric ansatz} and
\begin{align}
\hspace*{-3mm}h=f=1-\frac{2m}{r}-\frac{\beta^2}{2\eta r^2}-\frac{\Lambda r^2}{3}, \quad d\phi=\frac{\sqrt{2}\beta}{\eta r^2 \sqrt{h}} dr.
\end{align}
The horizon is sitting at $h=0$, 
\begin{equation}
\rh-{2m}-\frac{\beta^2}{2\eta \rh}-\frac{\Lambda \rh^3}{3}=0.
\end{equation}
The mass and surface gravity for this solution are
\begin{equation}
M= \frac{m}{\GN}, \qquad \kappa=\frac{\beta^2+2\eta (r_{_\text{H}}^2-\Lambda \rh^4)}{4\eta r^3_{_\text{H}}}.
\end{equation}
For this case \eqref{T-desired} becomes, 
\begin{equation}\label{Tbh-example-2}
    \th=T_0= \frac{\kappa}{2\pi}.
\end{equation}
For the entropy we need to apply the SPSM formula \eqref{entropy-formula}, which together with \eqref{Tbh-example-2}, after lengthy but straightforward algebra yields the area law again, $\Sbh=\pi \rh^2/\GN$. One can then readily observe that the first law $\th\delta\Sbh=\delta M$ is also satisfied.\\

{\centerline{\textbf{Discussion and Outlook}}} \vskip 2mm\label{sec:6}

We discussed that due to the presence of non-vanishing ambiguities, Wald entropy formula \eqref{SWald} does not necessarily yield the correct entropy for black holes in Horndeski theories. This matches with the previous observations that Wald entropy does not yield a well-defined first law of thermodynamics for such black holes \cite{Feng:2015wvb}. Our main result here is  that the resolution is in assigning a new temperature to these black holes corresponding to the surface gravity for gravitons, which together with the entropy variation computed using SPSM formulation yields the correct first law. 

The black hole temperature, as one expects both from Hawking \cite{Hawking:1974sw} or Unruh \cite{Unruh-temp} analysis should be a quantity determined only by  near horizon geometry. In ordinary cases all different light species in the problem near the horizon see a similar geometry and move with the same speed, in accordance with Einstein's equivalence principle. The key point in our analysis is that gravitons move on a different metric than the one the matter fields see, \eqref{Disformal-map}, as in many other beyond Einstein gravity theories see e.g. \cite{Choquet-Bruhat:2009xil, Izumi:2014loa}. Our results suggest that the relevant geometry for black hole thermodynamics is the one seen by gravitons. Among other things, this will provide a resolution to the species problem \cite{Species-1, Species-2}. 

That the effective metric for gravitons is the one relevant to black hole temperature and thermodynamics, may be checked further by repeating in more detail Hawking's process for these black holes and/or analyzing the Euclidean on-shell action \cite{Gibbons:1976ue} for our black hole solutions. The $\kappa$-peeling argument like those carried out in \cite{Peeling-2013} for gravitons suggests that Hawking analysis should yield the same temperature as ours in \eqref{T-desired}. Carrying out these analysis more closely could be illuminating. 

Due to presence of a profile of the scalar field we have a spontaneous breaking of Lorentz symmetry in the near horizon geometry, and this yields direction-dependent speed of gravitons \eqref{speed-of-graviton}. A similar feature is also shared by Einstein-Æther theories \cite{EE-theories}. It would hence be interesting to apply our ideas and results here to this framework and verify if they resolve similar issues about the black hole thermodynamics in those cases \cite{Eather-1, Eather-2}.

Assigning a temperature different than $\kappa/(2\pi)$ may raise the question about generalized second law of thermodynamics \cite{Bekenstein:1973ur}. Consider lump of  gas of photons of energy $\delta E\ll m_{\text{BH}}$ at temperature $T_\gamma$ falling into the hole. The first law implies  $\th\Delta \Sbh=\delta E=\frac34 T_\gamma S_{\gamma}$, where $\Delta \Sbh$ is the change in the entropy of the  hole  due to the fall and $S_{\gamma}$ is the entropy of the photon lump. The second law then requires, $\Delta \Sbh\geq S_\gamma$ or $\th\leq \frac34 T_\gamma$. For a photon to be absorbed into the hole  its wave-length should be smaller than $4\rh$ and hence $T_\gamma\sim (4\rh)^{-1}$. Therefore, a sufficient but not necessary condition for the second law is $\th \leq \frac{3}{16\rh}$, which is satisfied for our examples. 

Finally, we note that with the temperature \eqref{T-desired} in hand, one can fix the ambiguity in the Wald entropy analysis and provide a refined Wald entropy formula \cite{Work-in-progress}. \\

\begin{acknowledgments}
\textbf{Acknowledgments:} 
We are grateful to Gary Gibbons, Harvey Reall and Bayram Tekin for comments.
KH and MMShJ  thank the hospitality of ICTP HECAP  where this research carried out. MMShJ acknowledges the support by %grants from ICTP NT-04, 
INSF grant No 950124 and Saramadan grant No. ISEF/M/98204. KH acknowledges support from T$\ddot{\text{U}}$BITAK international researchers program 2221. SL acknowledges funding from the Ministry of Education and  Scientific Research (MIUR)  under the grant  PRIN MIUR 2017-MB8AEZ.

\end{acknowledgments}

\vskip 5mm

\centerline{\large{\textbf{Supplemental Material}}}

%\appendix
\vskip 3mm
\centerline{\textbf{Three more examples}}
\vskip 1mm

\noindent{\textbf{Example 3:}}
As our third example we analyze a 4-dimensional black brane with AdS$_4$ asymptotics which is a solution of the Lagrangian
\begin{equation}\label{L ex3}
\mathcal{L}=\frac{1}{16\pi\GN}\Big(R-2\Lambda-F_{\mu\nu}F^{\mu\nu}-2(\alpha g^{\mu\nu}-\gamma G^{\mu\nu})\partial_\mu\phi\partial_\nu\phi\Big)
\end{equation}
with arbitrary $\Lambda$ and $\alpha$. Comparing with Lagrangian \eqref{Horn-Lag}, we find $\cG_2=4\alpha\bx-2\Lambda$ and  $\cG=1+2\gamma\bx$. Instead of $\Lambda, \alpha$, one can use two other constants $\ell, \beta$  \cite{Feng:2015wvb}
\begin{equation}
\Lambda=-\frac{3(1+\frac{\beta}{2})}{\ell^2}, \qquad \alpha=\frac{3\gamma}{\ell^2}.
\end{equation}
In this convention, an electrically charged black brane solution in the coordinate $(t,r,x,y)$ is 
\begin{align}
&ds^2=-h(r)dt^2+\frac{dr^2}{f(r)}+r^2(dx^2+ dy^2),
\end{align}
where
\begin{align}
&h=\frac{r^2}{\ell^2}-\frac{m}{r}+\frac{4q^2}{(4+\beta)r^2}-\frac{4q^4\ell^2}{15(4+\beta)^2r^6}, \nonumber\\
&f=\frac{(4+\beta)^2r^8 h}{\big(\frac{2q^2\ell^2}{3}-(4+\beta)r^4\big)^2}, \nonumber\\
& d\phi=\sqrt{\frac{\beta-\frac{2q^2\ell^2}{3r^4}}{4\gamma f}}d r, \quad A=\big(\frac{q}{r}-\frac{2q^3\ell^2}{15(4+\beta) r^5} \big) dt.
\end{align}
The brane is situated at $\rh$ where $f(r_{_\text{H}})=h(r_{_\text{H}})=0$. Mass and electric charge ``densities" for this solution are
\begin{equation}
M=\frac{(4+\beta)m}{32\pi\GN}, \qquad Q=\frac{q}{4\pi\GN}.
\end{equation}
By densities it is understood that the charges are calculated without performing the integration over the $x$ and $y$ coordinates. Horizon surface gravity and electric potential are 
\begin{equation}
\kappa=\frac{3r_{_\text{H}}}{2\ell^2}-\frac{q^2}{(4+\beta)r^3_{_\text{H}}}, \qquad \Phi_{_\text{H}}=\frac{q}{r_{_\text{H}}}-\frac{2q^3\ell^2}{15(4+\beta) r^5_{_\text{H}}}.
\end{equation}

Insisting on the Hawking temperature $T_0=\frac{\kappa}{2\pi}$, the first law is not satisfied and the charge of the vector $\frac{1}{T_0}\xi_{_\text{H}}$ is not integrable \cite{Feng:2015wvb}. On the other hand, the new temperature in \eqref{T-desired} can be calculated to be 
\begin{equation}
T_{_{\text{\tiny{graviton}}}} = \left(\frac{3(4 +\beta)\rh ^4-2q^2\ell^2 }{12 r^4_{_\text{H}}}\right) T_0.
\end{equation}
This is exactly the $\th$ which makes the entropy integrable. The entropy can be calculated using SPSM to be found the Bekenstein-Hawking entropy density $\rh^2/(4\GN)$. It is easy to verify that the first law is also satisfied by the charge densities as
$$
\th\delta \Sbh =\delta M-\Phi_{_\text H}\delta Q.
$$

\noindent{\textbf{Example 4:}} As the fourth example, we study a rotating neutral BTZ-like black hole in 3 dimensional space-times which is a solution to the Lagrangian \eqref{L ex3}, and it is \cite{Santos:2020xox, BTZ-Horndeski}
\begin{align}
& ds^2=-h dt^2+\frac{dr^2}{h}+r^2 (d\varphi-\frac{j}{r^2}dt)^2, \nonumber\\ 
& h=-m+\frac{\alpha r^2}{\gamma}+\frac{j^2}{r^2}, \quad  d\phi=\sqrt{\frac{-(\alpha+\gamma\Lambda)}{2\alpha \gamma h}}  \ dr
\end{align}
where $\gamma <0$ and $(m,j)$ are free parameters in the solution. Mass,  angular momentum, horizon angular velocity, surface gravity, and horizon radii for this solution are
\begin{align}
&M=\frac{(\alpha -\Lambda \gamma)m}{16 \alpha \GN}, \qquad J=   \frac{(\alpha -\Lambda \gamma)j}{8 \alpha \GN},\nonumber\\
& \kappa_\pm=\frac{\alpha(r^2_+-r^2_-)}{\gamma r_\pm}, \qquad \Omega_\pm=\frac{j}{r_\pm^2},  \nonumber\\
& r_\pm^2=\frac{\gamma m \mp \sqrt{\gamma^2m^2-4\gamma\alpha j^2}}{2\alpha}. 
\end{align}
Notice that $\alpha<0$ in order to have positive horizon radii. By the new temperature in \eqref{T-desired}, one finds that
\begin{equation}
\th=\left(\frac{\alpha - \Lambda\gamma}{2\alpha}\right) T_0,     
\end{equation}
where $T_0=\frac{\kappa}{2\pi}$. Using this, \eqref{entropy-formula} yields $\Sbh=2\pi \rh/(4\GN)$ as the entropy of this black hole which satisfied the first law for each one of the horizons
$$
\th\delta \Sbh =\delta M-\Omega_{_\text H}\delta J.
$$

\noindent{\textbf{Example 5:}} The last example we present is a spherically symmetric neutral black hole solution of the  Horndeski theory \eqref{L ex3}. In 4-dimensional space-time the black hole solution is in the form of \eqref{metric ansatz} in which \cite{Rinaldi:2012vy}
\begin{align}
& h=1-\frac{2m}{r}+\frac{\alpha(4\alpha-\lambda)}{3\gamma(4\alpha+\lambda)}r^2+\frac{\lambda^2\sqrt{\frac{\gamma}{\alpha}}}{(16\alpha^2-\lambda^2)r}\tan^{-1}(\frac{r}{\sqrt{\frac{\gamma}{\alpha}}}), \nonumber\\ & f=\frac{(\gamma+\alpha r^2)h}{\gamma(r h)'}, \qquad d\phi=\sqrt{\frac{-\lambda(r^2h^2)' r}{8(\gamma+\alpha r^2)^2 h^2}}dr.
\end{align}
where $\lambda=2\alpha+2\gamma\Lambda$.  For this solution, the mass $M$ and the surface gravity $\kappa$ are equal to
\begin{equation}
M=\frac{\sqrt{16 \alpha^2-\lambda^2}}{4\alpha \GN}m, \ \kappa=\frac{\alpha(4\gamma+(4\alpha-\lambda) r_{_\text{H}}^2)}{2\gamma r_{_\text{H}}\sqrt{16\alpha^2-\lambda^2}},  
\end{equation}
where $r_{_\text{H}}$ is the root of $h(r)$.  
The Hawking temperature $T_0=\frac{\kappa}{2\pi}$ has the same problems as the other examples in this paper, i.e. the first law is not satisfied and entropy is not an integrable charge. The new temperature  \eqref{T-desired},
\begin{align}
 \th =T_{_{\text{\tiny{graviton}}}}= \left(\frac{4\gamma+(4\alpha-\lambda) r_{_\text{H}}^2}{4(\gamma+\alpha r_{_\text{H}}^2 )}\right) T_0,
\end{align}
resolves these problems and reproduces the Bekenstein-Hawking entropy  $\Sbh=\pi r_{_\text{H}}^2/\GN$, which satisfies the first law.
It is worth mentioning that in the presence of $\Lambda$, the first law can be extended to  include a Volume-Pressure term. One can find out \cite{Bayram} how this extension is possible by considering the $\Lambda$ as a conserved charge associated with a global gauge transformation \cite{Chernyavsky:2017xwm}. The bottom-line is that this extension is possible, and is compatible with the new temperature.\\

%\vskip 1mm
\centerline{\textbf{Details of ambiguity in Wald entropy}}
\vskip 2mm

Considering the extra term appearing in \eqref{X deformed} in Wald formula \eqref{SWald}, and using \eqref{binormal} for the Killing vector $\xih$, we find
\begin{align}
&\!\!\!\frac{\delta\big(R^{\rho\sigma}\partial_\rho\phi\partial_\sigma\phi \big)}{\delta R_{\alpha\beta\mu\nu}}\boldsymbol{\epsilon}_{\mu\nu}=\frac{\delta\big(R^{\rho\sigma}\partial_\rho\phi\partial_\sigma\phi \big)}{\delta R_{\alpha\beta\mu\nu}}\frac{\nabla_{[\mu}\xi_{\nu]}}{\kappa}\nonumber\\
&\!\!\!=(g^{\alpha\mu}\nabla^\beta\phi\nabla^{\nu}\phi)\frac{\nabla_{[\mu}\xi_{\nu]}}{\kappa}\nonumber\\
&\!\!\!=\frac{1}{\kappa}\Big(\nabla_{\mu}\Big((g^{\alpha\mu}\nabla ^{\beta}\phi\nabla ^{\nu}\phi)\xi_{\nu}\Big)-\nabla_{\mu}(g^{\alpha\mu}\nabla^{\beta}\phi\nabla^{\nu}\phi)\xi_{\nu}\Big)\nonumber\\
&\!\!\!=\frac{-1}{\kappa}\nabla^{\alpha}(\nabla^{\beta}\phi\nabla^{\nu}\phi)\xi_{\nu}=\frac{-1}{\kappa}\nabla^{\alpha}(\nabla^{\nu}\phi)\nabla^{\beta}\phi\xi_{\nu} \label{amb explicit}
\end{align}
in which we have used isometry condition $\xi_\mu\nabla^\mu\phi=0$. 

One may check that pull-back of the result to the bifurcation surface of horizon (which for our examples means choosing indices $(\alpha,\beta)=(t,r)$ and multiplying by $\sqrt{-g}$) is non-zero on the horizon. This contribution from \eqref{amb explicit} for the \textbf{Example 1} and for $\xih=\partial_t$ is 
\begin{equation}
\oint_{\text{H}}\frac{-\sqrt{-g}\, d\theta d \varphi }{\kappa} \nabla^{t}(\nabla^{\nu}\phi)\nabla^{r}\phi \, {\xih}_{\nu}=\frac{-\pi q^2}{\gamma}.    
\end{equation}
Therefore, there is a non-vanishing ambiguity in the Wald entropy.


\begin{thebibliography}{10}

\bibitem{Burgess-review}
C.~Burgess,
``Quantum gravity in everyday life: General relativity as an effective field theory,''
Living Rev. Rel. \textbf{7}, 5-56 (2004)
 \href {http://arxiv.org/abs/gr-qc/0311082}{[arXiv:gr-qc/0311082]}.

\bibitem{Clifton:2011jh}
T.~Clifton, P.~G.~Ferreira, A.~Padilla and C.~Skordis,
``Modified Gravity and Cosmology,''
Phys. Rept. \textbf{513} (2012), 1-189
\href {http://arxiv.org/abs/1106.2476}{[arXiv:1106.2476]}.

\bibitem{Nojiri}
S.~Nojiri and S.~D.~Odintsov,
``Modified f(R) gravity consistent with realistic cosmology: From matter dominated epoch to dark energy universe,''
Phys. Rev. D \textbf{74}, 086005 (2006)
\href {http://arxiv.org/abs/hep-th/0608008}{[arXiv:hep-th/0608008]}.

\bibitem{Creminelli:2018xsv}
P.~Creminelli and F.~Vernizzi,
``Dark Energy after GW170817 and GRB170817A,''
Phys. Rev. Lett. \textbf{119} (2017) no.25, 251302
\href {http://arxiv.org/abs/1710.05877}{[arXiv:1710.05877]}.

\bibitem{Sakstein:2017xjx}
J.~Sakstein and B.~Jain,
``Implications of the Neutron Star Merger GW170817 for Cosmological Scalar-Tensor Theories,''
Phys. Rev. Lett. \textbf{119} (2017) no.25, 251303
\href {http://arxiv.org/abs/1710.05893}{[arXiv:1710.05893]}.

\bibitem{Ezquiaga:2017ekz}
J.~M.~Ezquiaga and M.~Zumalacárregui,
``Dark Energy After GW170817: Dead Ends and the Road Ahead,''
Phys. Rev. Lett. \textbf{119} (2017) no.25, 251304
\href {http://arxiv.org/abs/1710.05901}{[arXiv:1710.05901]}.

\bibitem{Heisenberg:2018vsk}
L.~Heisenberg,
``A systematic approach to generalisations of General Relativity and their cosmological implications,''
Phys. Rept. \textbf{796} (2019), 1-113
\href {http://arxiv.org/abs/1807.01725}{[arXiv:1807.01725]}.

\bibitem{Barack:2018yly}
L.~Barack, et al, 
%V.~Cardoso, S.~Nissanke, T.~P.~Sotiriou, A.~Askar, C.~Belczynski, G.~Bertone, E.~Bon, D.~Blas, R.~Brito, T.~Bulik, C.~Burrage, C.~T.~Byrnes, C.~Caprini, M.~Chernyakova, P.~Chruściel, M.~Colpi, V.~Ferrari, D.~Gaggero, J.~Gair, J.~García-Bellido, S.~Hassan, L.~Heisenberg, M.~Hendry, I.~S.~Heng, C.~Herdeiro, T.~Hinderer, A.~Horesh, B.~J.~Kavanagh, B.~Kocsis, M.~Kramer, A.~Le Tiec, C.~Mingarelli, G.~Nardini, G.~Nelemans, C.~Palenzuela, P.~Pani, A.~Perego, E.~K.~Porter, E.~M.~Rossi, P.~Schmidt, A.~Sesana, U.~Sperhake, A.~Stamerra, N.~Tamanini, T.~M.~Tauris, L.~Arturo Urena-López, F.~Vincent, M.~Volonteri, B.~Wardell, N.~Wex, K.~Yagi, T.~Abdelsalhin, M.~Á.~Aloy, P.~Amaro-Seoane, L.~Annulli, M.~Arca-Sedda, I.~Bah, E.~Barausse, E.~Barakovic, R.~Benkel, C.~L.~Bennett, L.~Bernard, S.~Bernuzzi, C.~P.~Berry, E.~Berti, M.~Bezares, J.~J.~Blanco-Pillado, J.~L.~Blázquez-Salcedo, M.~Bonetti, M.~Bošković, Z.~Bosnjak, K.~Bricman, B.~Brüegmann, P.~R.~Capelo, S.~Carloni, P.~Cerdá-Durán, C.~Charmousis, S.~Chaty, A.~Clerici, A.~Coates, M.~Colleoni, L.~G.~Collodel, G.~Compère, W.~Cook, I.~Cordero-Carrión, M.~Correia, A.~de la Cruz-Dombriz, V.~G.~Czinner, K.~Destounis, K.~Dialektopoulos, D.~Doneva, M.~Dotti, A.~Drew, C.~Eckner, J.~Edholm, R.~Emparan, R.~Erdem, M.~Ferreira, P.~G.~Ferreira, A.~Finch, J.~A.~Font, N.~Franchini, K.~Fransen, D.~Gal'tsov, A.~Ganguly, D.~Gerosa, K.~Glampedakis, A.~Gomboc, A.~Goobar, L.~Gualtieri, E.~Guendelman, F.~Haardt, T.~Harmark, F.~Hejda, T.~Hertog, S.~Hopper, S.~Husa, N.~Ihanec, T.~Ikeda, A.~Jaodand, P.~Jetzer, X.~Jimenez-Forteza, M.~Kamionkowski, D.~E.~Kaplan, S.~Kazantzidis, M.~Kimura, S.~Kobayashi, K.~Kokkotas, J.~Krolik, J.~Kunz, C.~Lämmerzahl, P.~Lasky, J.~P.~Lemos, J.~Levi Said, S.~Liberati, J.~Lopes, R.~Luna, Y.~Z.~Ma, E.~Maggio, M.~M.~Montero, A.~Maselli, L.~Mayer, A.~Mazumdar, C.~Messenger, B.~Ménard, M.~Minamitsuji, C.~J.~Moore, D.~Mota, S.~Nampalliwar, A.~Nerozzi, D.~Nichols, E.~Nissimov, M.~Obergaulinger, N.~A.~Obers, R.~Oliveri, G.~Pappas, V.~Pasic, H.~Peiris, T.~Petrushevska, D.~Pollney, G.~Pratten, N.~Rakic, I.~Racz, F.~M.~Ramazanoğlu, A.~Ramos-Buades, G.~Raposo, M.~Rogatko, D.~Rosinska, S.~Rosswog, E.~Ruiz Morales, M.~Sakellariadou, N.~Sanchis-Gual, O.~S.~Salafia, A.~Sintes, M.~Smole, C.~Sopuerta, R.~Souza-Lima, M.~Stalevski, L.~C.~Stein, N.~Stergioulas, C.~Stevens, T.~Tamfal, A.~Torres-Forné, S.~Tsygankov, K.~v.~Ünlütürk, R.~Valiante, J.~Velhinho, Y.~Verbin, B.~Vercnocke, D.~Vernieri, R.~Vicente, V.~Vitagliano, A.~Weltman, B.~Whiting, A.~Williamson, H.~Witek, A.~Wojnar, K.~Yakut, H.~Yan, S.~Yazadjiev, G.~Zaharijas and M.~Zilhão,
``Black holes, gravitational waves and fundamental physics: a roadmap,''
Class. Quant. Grav. \textbf{36} (2019) no.14, 143001
%doi:10.1088/1361-6382/ab0587
\href {http://arxiv.org/abs/1806.05195}{[arXiv:1806.05195]}.

\bibitem{f(R)-review}
A.~De Felice and S.~Tsujikawa,
``f(R) theories,''
Living Rev. Rel. \textbf{13}, 3 (2010)
\href {http://arxiv.org/abs/1002.4928}{[arXiv:1002.4928]}.

T.~P.~Sotiriou and V.~Faraoni,
``f(R) Theories Of Gravity,''
Rev. Mod. Phys. \textbf{82}, 451-497 (2010)
\href {http://arxiv.org/abs/0805.1726}{[arXiv:0805.1726]}.

\bibitem{Lovelock}
D.~Lovelock,
``The Einstein tensor and its generalizations,''
J. Math. Phys. \textbf{12}, 498-501 (1971);
``The four-dimensionality of space and the einstein tensor,''
J. Math. Phys. \textbf{13}, 874-876 (1972)

\bibitem{Horndeski:1974wa} 
  G.~W.~Horndeski,
  ``Second-order scalar-tensor field equations in a four-dimensional space,''
  Int.\ J.\ Theor.\ Phys.\  {\bf 10}, 363 (1974).

\bibitem{Beyond-Horndeski-1}
D.~Langlois and K.~Noui,
``Degenerate higher derivative theories beyond Horndeski: evading the Ostrogradski instability,''
JCAP \textbf{02}, 034 (2016)
\href {http://arxiv.org/abs/1510.06930}{[arXiv:1510.06930]}.

\bibitem{Beyond-Horndeski-2}
J.~Gleyzes, D.~Langlois, F.~Piazza and F.~Vernizzi,
``Healthy theories beyond Horndeski,''
Phys. Rev. Lett. \textbf{114}, no.21, 211101 (2015)
\href {http://arxiv.org/abs/1404.6495}{[arXiv:1404.6495]}.

\bibitem{Beyond-Horndeski-3}
M.~Zumalacárregui and J.~Garc\'ia-Bellido,
``Transforming gravity: from derivative couplings to matter to second-order scalar-tensor theories beyond the Horndeski Lagrangian,''
Phys. Rev. D \textbf{89}, 064046 (2014)
\href {http://arxiv.org/abs/1308.4685}{[arXiv:1308.4685]}.

\bibitem{Beyond-Horndeski-4}
M.~Zumalacárregui and J.~Garc\'a-Bell
J.~Ben Achour, D.~Langlois and K.~Noui,
``Degenerate higher order scalar-tensor theories beyond Horndeski and disformal transformations,''
Phys. Rev. D \textbf{93}, no.12, 124005 (2016)
\href {http://arxiv.org/abs/1602.08398}{[arXiv:1602.08398]}.

\bibitem{Beyond-Horndeski-5}
T.~Kobayashi,
``Horndeski theory and beyond: a review,''
Rept. Prog. Phys. \textbf{82}, no.8, 086901 (2019)
\href {http://arxiv.org/abs/1901.07183}{[arXiv:1901.07183]}.

\bibitem{Abbott:2016blz} 
  B.~P.~Abbott {\it et al.} [LIGO Scientific and Virgo Collaborations],
  ``Observation of Gravitational Waves from a Binary Black Hole Merger,''
  Phys.\ Rev.\ Lett.\  {\bf 116}, no. 6, 061102 (2016). \href {http://arxiv.org/abs/1602.03837}{[arXiv:1602.03837]}.

\bibitem{LIGOScientific:2018mvr} 
  B.~P.~Abbott {\it et al.} [LIGO Scientific and Virgo Collaborations],
  ``GWTC-1: A Gravitational-Wave Transient Catalog of Compact Binary Mergers Observed by LIGO and Virgo during the First and Second Observing Runs,'' Phys. Rev. X \textbf{9} (2019) no.3, 031040
 \href {http://arxiv.org/abs/1811.12907}{[arXiv:1811.12907]}.
  
 \bibitem{Akiyama:2019cqa} 
  K.~Akiyama {\it et al.} [Event Horizon Telescope Collaboration],
 ``First M87 Event Horizon Telescope Results. I. The Shadow of the Supermassive Black Hole,''
  Astrophys.\ J.\  {\bf 875}, no. 1, L1 (2019). 
\href {http://arxiv.org/abs/1906.11238}{[arXiv:1906.11238]}.
  

\bibitem{Bardeen:1973gd}
J.M. Bardeen, B.~Carter, and S.W. Hawking, ``{The Four laws of black hole
	mechanics}'', {\em Commun.Math.Phys.}, {\textbf{31},\, }161--170, (1973).
	
\bibitem{Hawking:1974sw}
S.~Hawking,
``Particle Creation by Black Holes,''
Commun. Math. Phys. \textbf{43} (1975), 199-220. 

\bibitem{Bekenstein:1973ur}
J.~D.~Bekenstein,
``Black holes and entropy,''
Phys. Rev. D \textbf{7} (1973), 2333-2346.

\bibitem{Arnowitt:1960es}
R.~L. Arnowitt, S.~Deser, and C.~W. Misner, ``{Canonical variables for general
  relativity}'', {\em Phys. Rev.}, {\textbf{117},\, }1595--1602, (1960).

\bibitem{Arnowitt:1962hi}
R.~L. Arnowitt, S.~Deser, and C.~W. Misner, ``{The Dynamics of general
  relativity}'', {\em Gen. Rel. Grav.}, {\textbf{40},\, }1997--2027, (2008),
  \href {http://arxiv.org/abs/gr-qc/0405109} {[arXiv:gr-qc/0405109]}.

\bibitem{Wald:1993nt}
R.~M. Wald, 
``{Black hole entropy is the Noether charge}'', 
{\em Phys. Rev. D},
  {\textbf{48},\, }3427--3431, (1993), \href
  {http://arxiv.org/abs/gr-qc/9307038} {[arXiv:gr-qc/9307038]}.

\bibitem{Iyer:1994ys}
V.~Iyer and R.~M. Wald, 
``{Some properties of Noether charge and a proposal for  dynamical black hole entropy}'', 
{\em Phys. Rev. D}, {\textbf{50},\,
  }846--864, (1994), \href {http://arxiv.org/abs/gr-qc/9403028}
  {[arXiv:gr-qc/9403028]}.
  
 \bibitem{Feng:2015wvb} 
  X.~H.~Feng, H.~S.~Liu, H.~Lü and C.~N.~Pope,
  ``Thermodynamics of Charged Black Holes in Einstein-Horndeski-Maxwell Theory,''
  Phys.\ Rev.\ D {\bf 93}, no. 4, 044030 (2016)
  \href {http://arxiv.org/abs/1512.02659}{[arXiv:1512.02659]}. 

\bibitem{Gibbons:1996af} 
  G.~W.~Gibbons, R.~Kallosh and B.~Kol,
 ``Moduli, scalar charges, and the first law of black hole thermodynamics,''
  Phys.\ Rev.\ Lett.\  {\bf 77}, 4992 (1996)
  \href {http://arxiv.org/abs/hep-th/9607108}{[arXiv:hep-th/9607108]}.

\bibitem{Hajian:2016iyp} 
  K.~Hajian and M.~M.~Sheikh-Jabbari,
  ``Redundant and Physical Black Hole Parameters: Is there an independent physical dilaton charge?,''
  Phys.\ Lett.\ B {\bf 768}, 228 (2017)
 \href {http://arxiv.org/abs/1612.09279}{[arXiv:1612.09279]}.

\bibitem{Ashtekar:1987hia}
A.~Ashtekar, L.~Bombelli, and R.~Koul, 
``{Phase space formulation of general  relativity without a 3+1 splitting}'', 
{\em Lect. Notes Phys.},
  {\textbf{278},\, }356--359, (1987).

\bibitem{Ashtekar:1990gc}
A.~Ashtekar, L.~Bombelli, and O.~Reula, 
``{The covariant phase space of  asymptotically flat gravitational fields}'', 
{\em \emph{in M. Francaviglia
  (ed.),} Mechanics, Analysis and Geometry: 200 Years after Lagrange}, 417-450,
  (1990).

%\bibitem{Crnkovic:1987at}
%C.~Crnkovic and E.~Witten, 
%``{Covariant Description Of Canonical Formalism  In Geometrical Theories}'', 
%{\em \emph{In Hawking, S.W. (ed.), Israel, W.
%  (ed.):} Three hundred years of gravitation}, 676-684, (1987).

\bibitem{Lee:1990gr}
J.~Lee and R.~M. Wald, 
``{Local symmetries and constraints}'', 
{\em J. Math.  Phys.}, {\textbf{31},\, }725--743, (1990).

\bibitem{Barnich-Compere}
G.~Barnich and G.~Compere,
``Surface charge algebra in gauge theories and thermodynamic integrability,''
J. Math. Phys. \textbf{49} (2008), 042901
\href {http://arxiv.org/abs/0708.2378}{[arXiv:0708.2378]}.

\bibitem{Hajian:2015xlp} 
  K.~Hajian and M.~M.~Sheikh-Jabbari,
  ``Solution Phase Space and Conserved Charges: A General Formulation for Charges Associated with Exact Symmetries,''
  Phys.\ Rev.\ D {\bf 93}, no. 4, 044074 (2016)
 \href {http://arxiv.org/abs/1512.05584}{[arXiv:1512.05584]}.

\bibitem{Bettoni:2016mij}
D.~Bettoni, J.~M.~Ezquiaga, K.~Hinterbichler and M.~Zumalacárregui,
``Speed of Gravitational Waves and the Fate of Scalar-Tensor Gravity,''
Phys. Rev. D \textbf{95} (2017) no.8, 084029
\href {http://arxiv.org/abs/1608.01982}{[arXiv:1608.01982]}.

\bibitem{Maselli:2015yva}
A.~Maselli, H.~O.~Silva, M.~Minamitsuji and E.~Berti,
``Slowly rotating black hole solutions in Horndeski gravity,''
Phys. Rev. D \textbf{92}, no.10, 104049 (2015)
\href {http://arxiv.org/abs/1508.03044}{[arXiv:1508.03044]}.

\bibitem{Gleyzes:2013ooa}
J.~Gleyzes, D.~Langlois, F.~Piazza and F.~Vernizzi,
``Essential Building Blocks of Dark Energy,''
JCAP \textbf{08}, 025 (2013)
\href {http://arxiv.org/abs/1304.4840}{[arXiv:1304.4840]}.

\bibitem{Kovacs:2020ywu}
A.~D.~Kovacs and H.~S.~Reall,
``Well-posed formulation of Lovelock and Horndeski theories,''
\href {http://arxiv.org/abs/2003.08398}{[arXiv:2003.08398 [gr-qc]]}.

\bibitem{Kobayashi:2012kh}
T.~Kobayashi, H.~Motohashi and T.~Suyama,
``Black hole perturbation in the most general scalar-tensor theory with second-order field equations I: the odd-parity sector,''
Phys. Rev. D \textbf{85} (2012), 084025
\href {http://arxiv.org/abs/1202.4893}{[arXiv:1202.4893]}.

\bibitem{Kobayashi:2014wsa}
T.~Kobayashi, H.~Motohashi and T.~Suyama,
``Black hole perturbation in the most general scalar-tensor theory with second-order field equations II: the even-parity sector,''
Phys. Rev. D \textbf{89} (2014) no.8, 084042
\href {http://arxiv.org/abs/1402.6740}{[arXiv:1402.6740]}.

\bibitem{Work-in-progress}
K. Hajian, S. Liberati, M.M. Sheikh-Jabbari, M.H. Vahidinia, ``{On entropy and thermodynamics of black holes in Horndeski theories,''},  \emph{In preparation.}

\bibitem{Ghodrati:2016vvf} 
  M.~Ghodrati, K.~Hajian and M.~R.~Setare,
  ``Revisiting Conserved Charges in Higher Curvature Gravitational Theories,''
  Eur.\ Phys.\ J.\ C {\bf 76}, no. 12, 701 (2016)
 \href {http://arxiv.org/abs/1606.04353}{[arXiv:1606.04353]}.

\bibitem{Hajian:2015eha} 
  K.~Hajian, Ph.D thesis,
  ``On Thermodynamics and Phase Space of Near Horizon Extremal Geometries,''
  \href {http://arxiv.org/abs/1508.03494}{[arXiv:1508.03494]}.

\bibitem{Seraj:2016cym} 
  A.~Seraj, Ph.D thesis,
  ``Conserved charges, surface degrees of freedom, and black hole entropy,''
\href {http://arxiv.org/abs/1603.02442}{[arXiv:1603.02442]}.

\bibitem{Compere-lecture}
G.~Compère and A.~Fiorucci,
``Advanced Lectures on General Relativity,''
\href {http://arxiv.org/abs/1801.07064}{[arXiv:1801.07064]}.

\bibitem{Bettoni:2013diz}
D.~Bettoni and S.~Liberati,
``Disformal invariance of second order scalar-tensor theories: Framing the Horndeski action,''
Phys. Rev. D \textbf{88} (2013), 084020
\href {http://arxiv.org/abs/1306.6724}{[arXiv:1306.6724]}.

\bibitem{Peeling-2013}
B.~Cropp, S.~Liberati and M.~Visser,
``Surface gravities for non-Killing horizons,''
Class. Quant. Grav. \textbf{30} (2013), 125001
%doi:10.1088/0264-9381/30/12/125001
\href {http://arxiv.org/abs/1302.2383}{[arXiv:1302.2383]}.


  \bibitem{Cisterna:2014nua} 
  A.~Cisterna and C.~Erices,
  ``Asymptotically locally AdS and flat black holes in the presence of an electric field in the Horndeski scenario,''
  Phys.\ Rev.\ D {\bf 89}, 084038 (2014)
 \href {http://arxiv.org/abs/1401.4479}{[arXiv:1401.4479]}.

 \bibitem{Miao:2016aol} 
  Y.~G.~Miao and Z.~M.~Xu,
  ``Thermodynamics of Horndeski black holes with non-minimal derivative coupling,''
  Eur.\ Phys.\ J.\ C {\bf 76}, no. 11, 638 (2016)
 \href {http://arxiv.org/abs/1607.06629}{[arXiv:1607.06629]}.

\bibitem{Hu:2018qsy} 
  Y.~P.~Hu, H.~A.~Zeng, Z.~M.~Jiang and H.~Zhang,
  ``P-V criticality in the extended phase space of black holes in Einstein-Horndeski gravity,''
  Phys.\ Rev.\ D {\bf 100}, no. 8, 084004 (2019)
  \href {http://arxiv.org/abs/1812.09938}{[arXiv:1812.09938]}.


\bibitem{Babichev:2017guv}
E.~Babichev, C.~Charmousis and A.~Lehébel,
``Asymptotically flat black holes in Horndeski theory and beyond,''
JCAP \textbf{04} (2017), 027
\href {http://arxiv.org/abs/1702.01938}{[arXiv:1702.01938]}.

\bibitem{Unruh-temp}
W.~Unruh,
``Notes on black hole evaporation,''
Phys. Rev. D \textbf{14} (1976), 870.
%doi:10.1103/PhysRevD.14.870

%
\bibitem{Choquet-Bruhat:2009xil}
Y.~Choquet-Bruhat,
``The Cauchy Problem for Stringy Gravity,''
J. Math. Phys. \textbf{29} (1988), 1891-1895; 
%doi:10.1063/1.527841Y.~Choquet-Bruhat,
``General Relativity and the Einstein Equations,'' \href{https://books.google.it/books?id=QkUTDAAAQBAJ&lpg=PP1&pg=PR21#v=onepage&q&f=false}{\textit{Oxford Mathematical Monographs,} 2008.} 

\bibitem{Izumi:2014loa}
K.~Izumi,
``Causal Structures in Gauss-Bonnet gravity,''
Phys. Rev. D \textbf{90} (2014) no.4, 044037, 
%doi:10.1103/PhysRevD.90.044037
\href {http://arxiv.org/abs/1406.0677}{[arXiv:1406.0677]}.


\bibitem{Species-1}
L.~Susskind and J.~Uglum,
``Black hole entropy in canonical quantum gravity and superstring theory,''
Phys. Rev. D \textbf{50} (1994), 2700-2711
%doi:10.1103/PhysRevD.50.2700
\href {http://arxiv.org/abs/94010709}{[arXiv:hep-th/94010709]}.

%\cite{Susskind:2005js}
%\bibitem{Susskind:2005js}
L.~Susskind and J.~Lindesay,
``An introduction to black holes, information and the string theory revolution: The holographic universe,'' World Scientific (2005) 183pp.

\bibitem{Species-2}
T.~Jacobson,
``Black hole entropy and induced gravity,''
\href {http://arxiv.org/abs/9404039}{[arXiv:gr-qc/9404039]}.

\bibitem{Gibbons:1976ue}
G.~Gibbons and S.~Hawking,
``Action Integrals and Partition Functions in Quantum Gravity,''
Phys. Rev. D \textbf{15} (1977), 2752-2756.
%doi:10.1103/PhysRevD.15.2752

\bibitem{EE-theories}
B.~Cropp, S.~Liberati, A.~Mohd and M.~Visser,
``Ray tracing Einstein-Æther black holes: Universal versus Killing horizons,''
Phys. Rev. D \textbf{89} (2014) no.6, 064061
%doi:10.1103/PhysRevD.89.064061
 \href {http://arxiv.org/abs/1312.0405}{[arXiv:1312.0405]}.


\bibitem{Eather-1}
P.~Berglund, J.~Bhattacharyya and D.~Mattingly,
``Towards Thermodynamics of Universal Horizons in Einstein-æther Theory,''
Phys. Rev. Lett. \textbf{110} (2013) no.7, 071301
%doi:10.1103/PhysRevLett.110.071301
\href {http://arxiv.org/abs/1210.4940}{[arXiv:1210.4940]}.


\bibitem{Eather-2}
C.~Pacilio and S.~Liberati,
``Improved derivation of the Smarr formula for Lorentz-breaking gravity,''
Phys. Rev. D \textbf{95} (2017) no.12, 124010
%doi:10.1103/PhysRevD.95.124010
\href {http://arxiv.org/abs/1701.04992}{[arXiv: 1701.04992]};
%\bibitem{Liberati:2017vse}
%C.~Pacilio and S.~Liberati,
``First law of black holes with a universal horizon,''
Phys. Rev. D \textbf{96} (2017) no.10, 104060
%doi:10.1103/PhysRevD.96.104060
\href {http://arxiv.org/abs/1709.05802}{[arXiv:1709.05802]}.

\bibitem{Santos:2020xox}
F.~F.~Santos,
``Rotating black hole with a probe string in Horndeski Gravity,''
Eur. Phys. J. Plus \textbf{135} (2020) no.10, 810
\href {http://arxiv.org/abs/2005.10983}{[arXiv:2005.10983]}.

\bibitem{BTZ-Horndeski}
M.Bravo-Gaete and M.Hassaine,
``Thermodynamics of a BTZ black hole solution with an Horndeski source,''
Phys. Rev. D 90 (2014) no.2, 024008, 
%doi:10.1103/PhysRevD.90.024008
\href {http://arxiv.org/abs/1405.4935}{[arXiv:1405.4935 [hep-th]]}.

%F.~Santos,
%``Rotating black hole with a probe string in Horndeski Gravity,''
% \href {http://arxiv.org/abs/2005.10983}{[arXiv:2005.10983]}.


\bibitem{Rinaldi:2012vy}
M.~Rinaldi,
``Black holes with non-minimal derivative coupling,''
Phys. Rev. D \textbf{86} (2012), 084048
\href {http://arxiv.org/abs/1208.0103}{[arXiv:1208.0103]}.

\bibitem{Bayram}
Kamal Hajian, Hikmet Ozsahin, Bayram Tekin, Work in progress.

\bibitem{Chernyavsky:2017xwm} 
  D.~Chernyavsky and K.~Hajian,
  ``Cosmological constant is a conserved charge,''
  Class.\ Quant.\ Grav.\  {\bf 35}, no. 12, 125012 (2018)
 \href {http://arxiv.org/abs/1710.07904}{[arXiv:1710.07904]}.



\end{thebibliography}
\end{document}